# Asymmetric Coulomb Oscillation and Giant Anisotropic Magnetoresistance in Doped Graphene Nanojunctions

Subramani Amutha[1,2] and Arijit Sen[1,2] *

*[1]SRM Research Institute, SRM University, Kattankulathur PO, Chennai-603203, India*

*[2]Department of Physics & Nanotechnology, Kattankulathur PO, Chennai-603203, India*

*Email: arijit.s@res.srmuniv.ac.in*

ABSTRACT

We report here the charge transport behavior in graphene nanojunctions in which graphene nanodots, with relatively long relaxation time, are interfaced with ferromagnetic electrodes. Subsequently we explore the effect of substitutional doping of transition metal atoms in zigzag graphene nanodots (z-GNDs) on the charge transport under non-collinear magnetization. Only substitutional doping of transition metal atoms in z-GNDs at certain sites demonstrates the spin filtering effect with a large tunnelling magnetoresistance as high as 700%, making it actually suitable for spintronic applications. From the electrical field simulation around the junction area within the electrostatic physics model, we find that the value of electric field strength increases especially with doped graphene nanodots, as the gap between the gate electrode and tip axis is reduced from 3 nm to 1 nm. Our detailed analysis further suggests the onset of asymmetric *Coulomb oscillations* with varying amplitudes in graphene nanodots, on being doped with magnetic ions. Such kind of tunability in the electronic conductance can potentially be exploited in designing spintronic logic gates at nanoscale.

## 1. INTRODUCTION

The fabrication of single and few layers graphene has sparked unending expectations in the field of all-carbon based applications ranging from electronics to nano-spintronics [1-6]. Graphene has recently been exploited for conductive switching and bio-imaging [7-8] as well. An array of graphene sheet that forms a one-dimensional (1D) graphene nanoribbon (GNR) is fascinating due to the physiochemical properties at nanoscale, especially the existence of edges and orders [9,10] as obtained out of unzipping carbon nanotubes (CNT) [11,12] and lithography [13,14]. GNR provides some remarkable physical properties to behave as a spin-filter [15-17] or a spin-valve [18-21]. It is broadly classified into two categories based on the geometry: armchair graphene nanoribbon (aGNR) and zigzag graphene nanoribbon (zGNR) [22,23]. Son *et al* [24] have shown that GNRs with zigzag edges may be created to hold a spin current within the presence of a sufficiently massive field of force, opening up avenues for the design of spin-dependent switching devices. Several recent studies have reported the half-metallicity [24] in zGNRs on chemical doping, whereas the applying of an external field  leads to trigger a transition from parallel to anti-parallel magnetic edges, resulting in large magnetoresistance [25] phenomena suitable for economical spintronic devices. A zigzag graphene nanoribbon of finite length behaves like a graphene nanodot (GND) with a fairly long time constant. As compared to zGNRs, the charge transport phenomena in zigzag graphene nanodots (z-GNDs) are rather less understood, especially when subject to non-collinear spin polarization.

For functionalization of z-GNDs, substitutional doping can play a crucial role. However, the selection of dopant and its position is also important for designing better nanodevices.  According to Santos *et al* [26], nickel doping in GND can give better results since nickel impurity entails strong covalent interaction with carbon. Interestingly, edges are the most favorable sites for Ni-doping [27] in such systems. Further, substitutional doping reveal enhanced metallicity for all the doping sites. Ushiro *et al* [28] have also shown that after careful purification in SWCNTs, the Ni substitutional defects can be such that the most likely configuration is the one in which the Ni atom replaces a carbon atom. The electronic and magnetic properties are drastically influenced by the presence of dopants, in the form of defects. The magnetoresistance in magnetic tunnel junctions may also depend on the orientation of magnetizations in the ferromagnetic electrodes which introduce a phenomenon called the tunneling anisotropic magnetoresistance (TAMR) effect. Recent experimental reports show a linear magnetoresistance of 80–250% in epitaxial multilayer graphene [29], and a quadratic magnetoresistance of 60%, at 300 K in few-layer graphene [30] grown by chemical vapor deposition (CVD).

In this work, we show that a small-sized graphene nanodot can also exhibit an extremely large TAMR, as we investigate Ni-interaction with Z-GNDs. Our focus remains on the spin-dependent charge transport characteristics across zigzag graphene nanodots interfaced with magnetic Ni-electrodes. We carry out this study within the framework of the density functional theory (DFT) combined with the non-equilibrium Green's function (NEGF) formalism.

## 2. METHODOLOGY

We optimized a set of Ni|z-GND|Ni device geometries by making use of DFT, as implemented in the SIESTA package [31]. However, the pre-optimized Ni-electrodes were kept fixed while relaxing the central part consisting of GNDs that form the respective molecular moieties. We utilized norm-conserving pseudopotentials to treat the atomic cores. For studying the spin-dependent charge transport properties, we initially computed the ground-state Hamiltonian using DFT. Subsequently, the quantum transport calculations were performed based on the non-equilibrium Green's function (NEGF) formalism [32] to take care of the nonequilibrium quantum statistics of the respective device. As in the case of geometry optimization, we employed the double-ς with polarization (DZP) basis set throughout our quantum transport simulations, based on the NEGF-DFT framework [33-35]. The density mesh cutoff was set to 200 Ry, while the $k$-point sampling grid remained as $1 \times 1 \times 400$.

From the self-consistent device Hamiltonian, the spin-resolved current can be obtained as

$$I_\sigma(V_b) = \frac{e}{h} \int dE\, T_\sigma(E, V_b)[f(E - \mu_L) - f(E - \mu_R)], \qquad (1)$$

where $\mu_L$ ($\mu_R$) denotes the chemical potential of the left (right) electrode, while $f(\dots)$ refers to the Fermi-Dirac distribution functions. On the other hand, the spin-polarized transmission takes the following form:

$$T_\sigma(E, V_b) = Tr[\Gamma_L(E - qV_b)G^r(E)\Gamma_R(E - qV_b)G^a(E)]_{\sigma\sigma} \quad (2)$$

Here, $\Gamma_L$ ($\Gamma_R$) signifies the coupling strengths between the left (right) semi-infinite ferromagnetic electrodes and the scattering region that includes a z-GND [36]. In addition, $G^r$ ($G^a$) defines the retarded (advanced) Green's function having a $2n$ x $2n$ NEGF matrix, $n$ being the size of the basis set.

## 3. RESULTS AND DISCUSSIONS

### 3.1. Spin-dependent transmission

Fig. 1(a) shows the model structure of Ni/z-GND/Ni junction. The model optimizes the junction of Ni doped z-GND with Ni magnetic electrodes. In order to obtain preferential doping positions of Ni atom in z-GND, we have calculated various doping sites which are shown in Fig.1. The carbon atoms at the edge of z-GND are passivated by hydrogen atoms in the pristine model. The Ni atom is embedded in the z-GND by replacing two C atoms at both edges. For different substitutional doping sites, we choose the one at the top edge of the z-GND to be only varied while keeping the other at bottom edge fixed, as shown in Fig. 1 (c-e). The undoped z-GNDs are denoted as $S_0$, whereas the doped nanostructures are referred to as $S_1$, $S_2$ and $S_3$ in this content. Fig 1(b) displays $S_0$, the undoped (*i.e.* pristine) z-GNDs where the transmission profile show most of the transmission in HOMO level from moderate to high, irrespective of the spin orientation. As the contour plots of Fig. 1 (b-e) suggest, the electronic transmission for $S_1$ is found to be strongly dependent on the spin orientation, whereas it is rather more sensitive to the electronic energy for $S_0$, $S_2$ and $S_3$. The latter part is more evident from the transmission profiles of Fig. 1 (f), where only the HOMO channels dominate for $S_0$, $S_2$ and $S_3$, in contrast to $S_1$. Ni-doped z-GNDs are known to be quite stable due to the high binding energy and also, for the energy barrier of Ni atoms [37]. From the observed transmission profiles, the transmission is strongly dependent on the position of the Nickel doping with z-GNDs implying the major role of the Ni-site in determining the angular dependence of transmission, conductance and hence, magnetoresistance.

### 3.2. Angular dependence of conductance

The angular dependence of electronic conductance is shown for pure and Ni-doped z-GNDs in Fig.2. As it appears, the high conductance (HC) for undoped GND nanojunctions, labeled as $S_0$, remains nearly constant with respect to the electrode spin orientations. A similar trend is also observed for Ni-doped z-GND nanojunctions, labeled as $S_3$, although at sufficiently low conductance (LC) values. Such features of binary conductance can be potentially exploited as a tunable ON/OFF switch at nanoscale. The two states of electronic conductance obtained by simply tuning the electrode magnetization may well serve as 1s and 0s of binary logic to find potential utilities in designing spintronic logic circuits at nanoscale.

By introducing initially the Ni-dopants to z-GNDs and then manipulating the positions of Ni at different sites, *viz.* $S_1$ and $S_2$, we come across a jump in the electronic conductance while being dependent highly on the spin-orientation (see Fig. 2). It turns out that Ni-doped z-GNDs with site position $S_1$ may attain higher conductance limit with respect to the angular variation of the electrode spin than those with site position $S_2$. For example, the $S_1$ nanojunction exhibits the electronic conductance as high as $1.3G_0$, where $G_0$ is the conductance quantum, for spin orientation of $\theta = \pi/2$ followed by an attenuation of $0.6G_0$, at $\theta = 0.6\pi$. As demonstrated by Fig. 2, the angular dependence of electronic conductance for $S_1$ and $S_2$ nanojunctions behaves in an asymmetrically oscillating manner, which may be accorded to the asymmetricity in the respective sites of the two Ni-dopants. Thus the electronic transport properties of z-GNDs can be modulated by simply tuning the spin orientation after manipulating the relative position of Ni dopant with respect to the other one being at the opposite edge of the graphene nanodot.

### 3.3 Angular dependence of magnetoresistance

Fig.3 shows the tunneling anisotropic magneto-resistance (*TAMR*) for a Ni|z-GND|Ni system, which may be expressed as

$$TAMR = \frac{R(\theta) - R(0)}{R(0)}, \tag{3}$$

where $R(\theta)$ denotes the device resistance as a function of the electrode spin orientation ($\theta$), while $R(0)$ is associated with $\theta$=0 [38]. It turns out that *TAMR* remains invariant under spin orientation for the undoped nanojunctions while for the doped nanojunctions, it has always a strong angular dependence (see the inset of Fig. 3). We find that there is a sudden jump in *TAMR* value of about 700% at $\theta = 0.9\pi$ for the $S_2$ nanojunctions where the dopant centers are coupled more asymmetrically. This is in good agreement with the occurrence of conductance jumps upon rotation of the magnetization, as reported in Ref. 18.

## 3.4 Electrostatic simulations

Simulation for the electrical field along with the potential distribution around various graphene nanojunctions was carried out by making use of the COMSOL Multiphysics tool using electrostatic physics model [39]. We chose a cone-like tip for each of the three electrodes (*viz.* source, drain and gate) while the nano-moiety in the central part was designed with an ellipsoidal shape. The distance between source and drain came out to be 1.3 nm, as estimated from the SIESTA optimization for each nanojunction. However, the gap between the as-introduced gate electrode and tip axis was varied from 1 nm to 3 nm with the molecular length being close to 1.1 nm. The source to drain voltage was kept fixed at 0.1 V while the gate voltage was set as -4 V. A proper selection of the gate potential was necessary for tuning the transport characteristics of various nanojunctions.

Our results reveal the potential distribution around the gate electrode to be inhomogeneous and are strongly dependent on the location of the gate electrodes with respect to the source and drain electrodes, as shown in Fig. 4. The position of the gate electrode is positioned at the center between the source and drain electrodes to extract the potential distribution from the simulation. In the case of $S_1$ and $S_2$, the potential distribution is high and it may reach maximum electric field of 0.3 x $10^{10}$ V/m and 0.26 x $10^{10}$ V/m respectively at a distance of 2 nm between the gate electrodes. For $S_0$ and $S_3$, the potential distribution becomes minimal with the electric field value of 0.24 x $10^{10}$ V/m. Further, the electrodes placed at a distance of 1 nm and 3 nm possess negligible electric field when compared to 2 nm for all the systems. Thus we see that positioning the gate electrode at 2 nm increases the electric field distribution around the molecular junction which may be helpful for efficient design of graphene nanodot based molecular transistors.

## 4. CONCLUSIONS

Banking on the *first-principles* quantum transport approach, we have investigated the anisotropic behavior in the charge transport as well as the tunnel magnetoresistance across doped graphene nanodots. We observe that even a small adjustment in the spin orientation of the magnetic electrodes may cause a rapid enhancement in the junction conductance. Further, asymmetric

Coulomb oscillation turns out to be maximal for Ni-doped graphene nanodots, as compared to the undoped one. Since, tunability in anisotropic magnetoresistance would be of great advantage in molecular switching, the present study can potentially offer some useful insights on how to exploit graphene nanodots in memory devices. We have also shown that it is possible to locate the particular position of the gate electrode so that the high-potential regions can be mapped across the terminals, which may further help in designing graphene nanodot based molecular transistors.

## ACKNOWLEDGEMENT

This work was supportedbyDST NanoMission, Govt. of India, *via* Project No. SR/NM/NS1062/2012. We are thankful to the National PARAM Supercomputing Facility (NPSF), Centre for Development of Advanced Computing (C-DAC), along with SRM-HPCC, for facilitating the high-performance computing.

**Figure captions:**

Fig.1: (a) Schematic of a doped zigzag graphene nanodot (z-GND) based spintronic nanodevice. (b-e) Contour plots of electron transmission in several nanojunctions comprising undoped as well as doped z-GNDs, being subject to different spin orientations. (f) Transmission profile for these nanojunctions, as a function of the electronic energy relative to the Fermi level.

Fig. 2 : Angular dependence of the electronic conductance for undoped as well as doped Ni|z-GND|Ni nanojunctions.

Fig. 3: Angular dependence of the tunnel anisotropic magnetoresistance (TAMR) for undoped as well as doped Ni|z-GND|Ni nanojunctions.

Fig.4: Electric field simulation and Potential distribution for undoped as well as doped Ni|z-GND|Ni nanojunctions.

# **Figures**

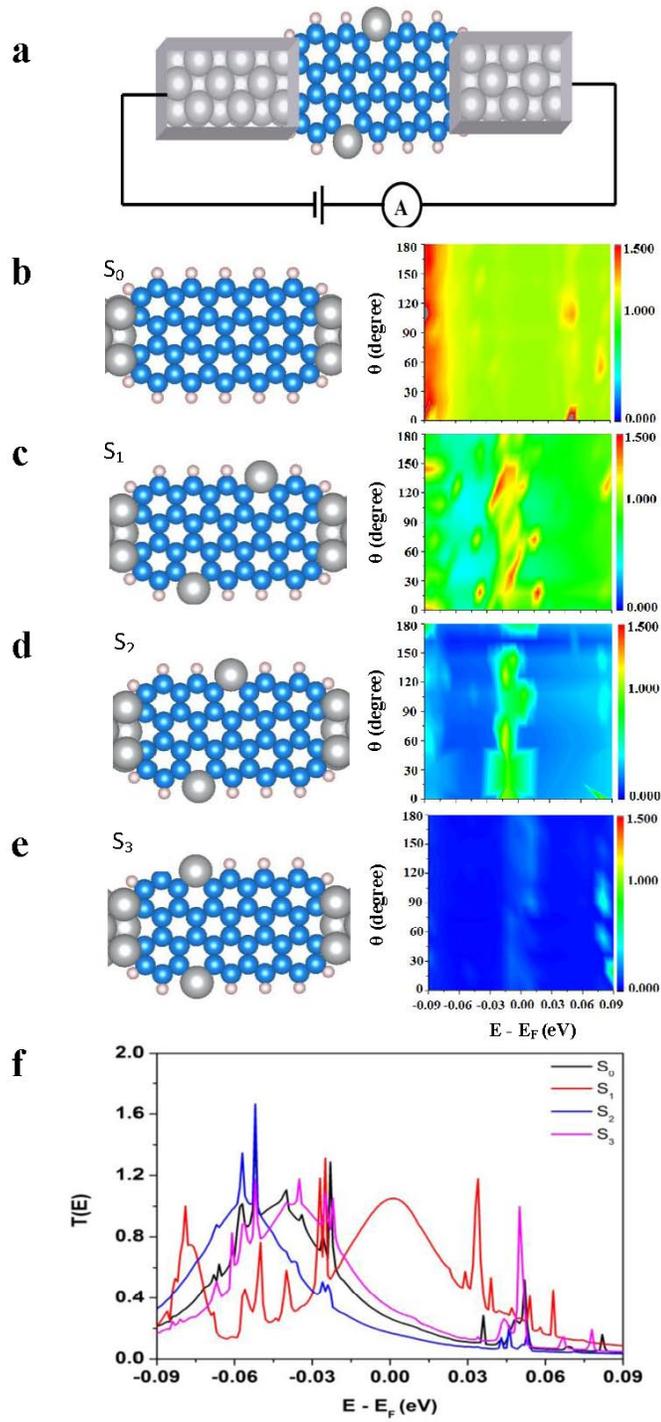

**Figure 1**

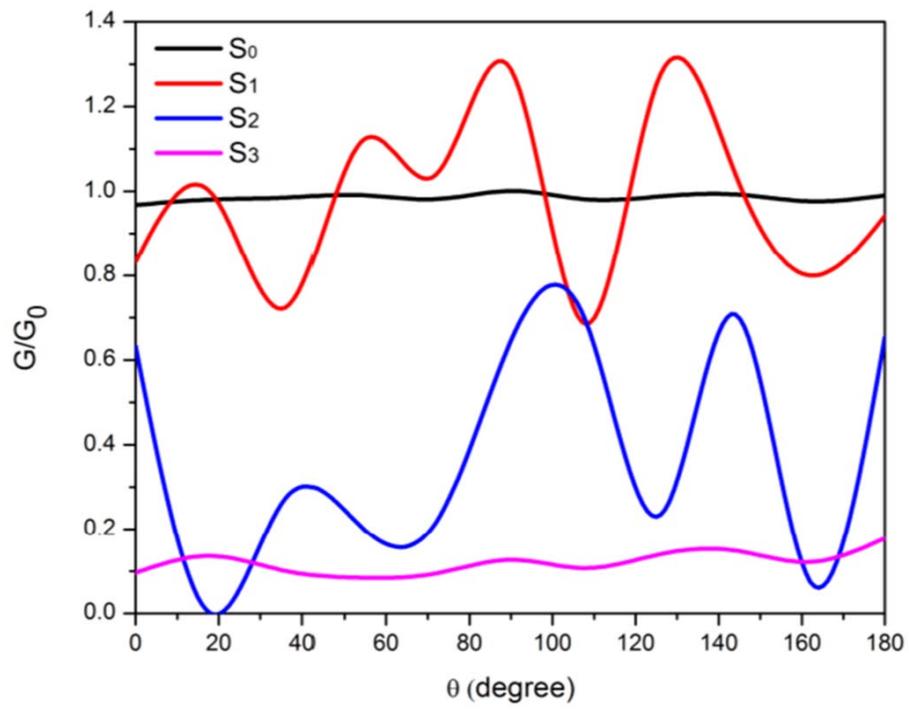

**Figure 2**

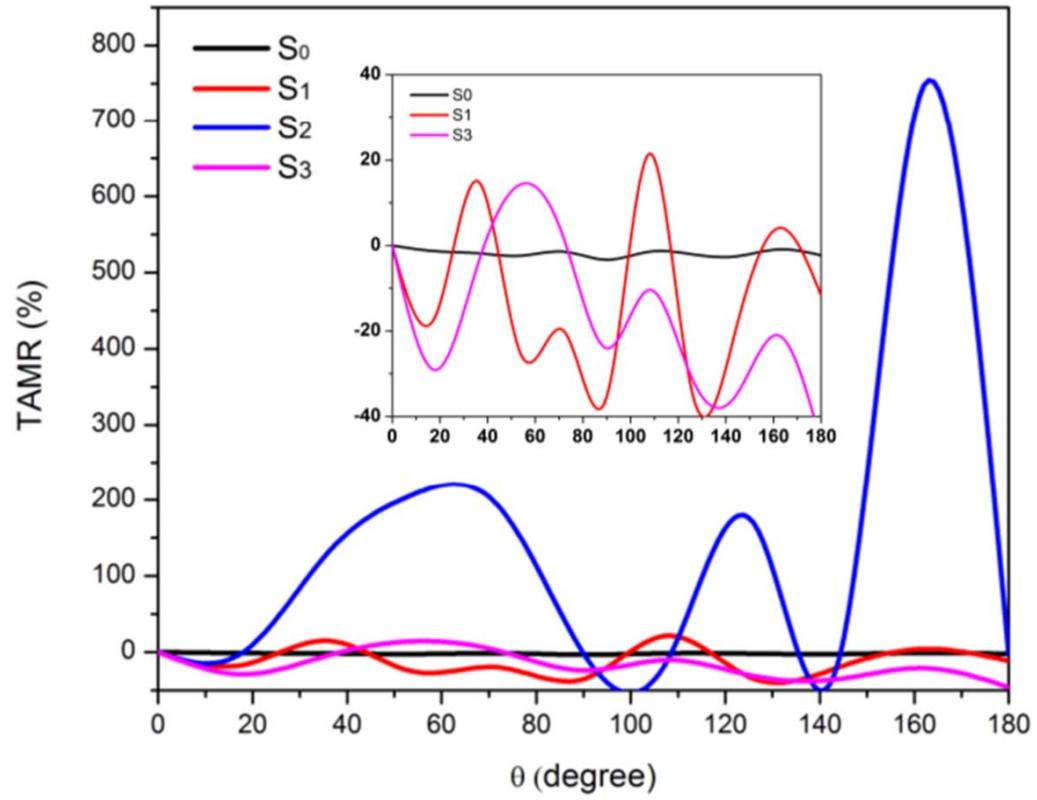

**Figure 3**

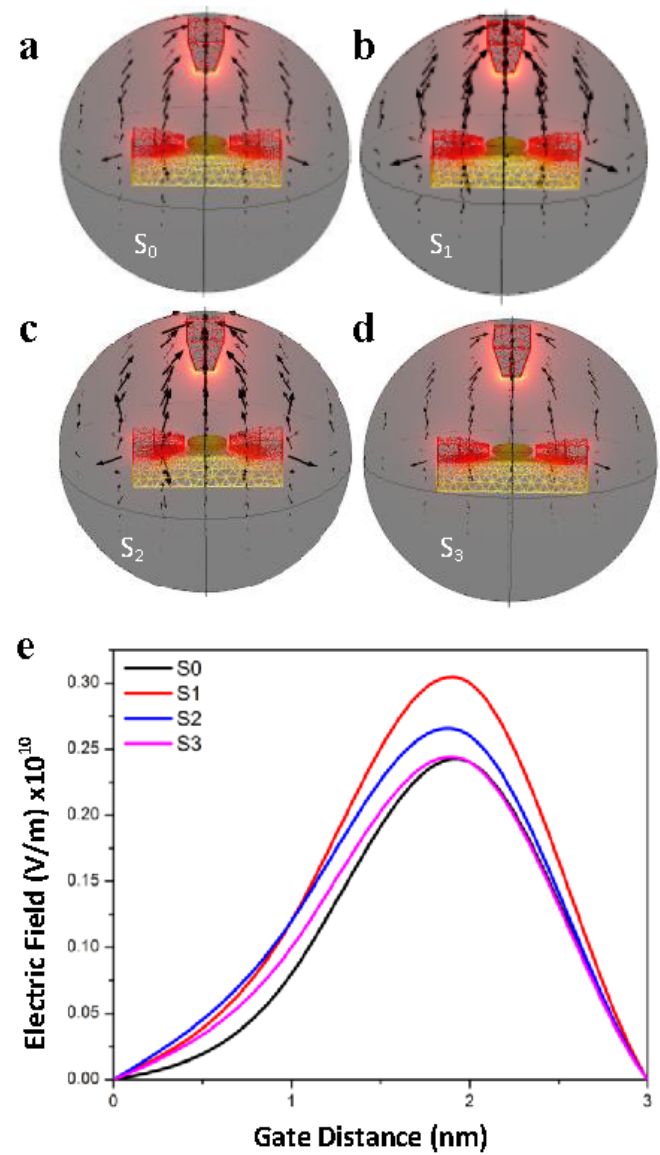

**Figure 4**